\documentstyle[preprint,floats,prd,aps]{revtex}

\begin{document}
\preprint{\vbox{\hbox{UH 511-828-95 \hfill}
                \hbox{February 25, 1995 \hfill}}}
\input psfig
\title{
Experimental Implications  of Large CP Violation and Final
State Interactions in
the Search for $D^0-\bar{D}^0$ Mixing.}
\author{ T.E. Browder and S. Pakvasa}
\address{University of Hawaii at Manoa, Honolulu, Hawaii 96822}
\maketitle
\begin{abstract}
We discuss the implications of CP violation
as well as final state interaction phases in the experimental
search for $D^0-\bar{D}^0$ mixing. At the present level
of sensitivity, these are not yet a 
significant systematic experimental limitation.
\end{abstract}
\bigskip

\section{Introduction}

As was recently noted by Blaylock, Seiden, and Nir\cite{Blaylock}
due to final state interaction (FSI) a term
proportional to $\Delta M ~t~ e^{-\Gamma t}$ may appear in the rate
of wrong sign $D$ decays 
even in the absence of CP violation. Moreover, in some extensions
of Standard Model which have large values of both $\Delta M$ 
and significant CP violation, a similar term may arise.
Blaylock {\it et al.} have suggested that a value of 
$\Delta M$ larger than the
present experimental limit can be accomodated if one of these 
previously neglected terms 
destructively interferes with the other time dependent terms
which arise from mixing (proportional to $t^2 ~e^{-\Gamma t}$)
 and from doubly Cabibbo suppressed decays (DCSD) (proportional 
to $e^{-\Gamma t}$). 
They suggest that this may invalidate the use of existing limits from time
dependent mixing studies at fixed target 
experiments \cite{E691} , \cite{E791}
to constrain extensions of the Standard Model.

Below, we give expressions for the time
dependence in the general case and then
attempt to estimate the 
maximum size of the terms proportional to $t e^{-\Gamma t}$.

\section{Formalism for Mixing}

We follow the notation of references
\cite{Blaylock},\cite{liu1},\cite{liu2}. Let the mass eigenstates
be $D_S$, $D_L$. Then 
\[
|D_S> = p |D^0> + q |\bar{D}^0>
\]
\[
|D_L> = p |D^0> - q |\bar{D}^0>
\]
In the limit of no CP violation, $p$=$q$=$1/\sqrt{2}$.

Let $\Delta M= M_L- M_S$ and $\Delta \Gamma = \Gamma_L-\Gamma_S$
denote the mass difference and lifetime difference, respectively.
Let A denote the amplitude for $<f|H|D^0>$, 
B the amplitude for $<f|H|\bar{D}^0>$. Let
$\displaystyle\lambda={p \over q}{A\over B}$ and
$\displaystyle\bar{\lambda}={q \over p}{\bar{A}\over \bar{B}}$.
The decay rate is then given by
\begin{equation}
\Gamma(D^0(t)\to K^+\pi^-) = {e^{-\Gamma t} \over 4} |B|^2 |{q\over p}|^2
\{ 4 |\lambda|^2 + ({\Delta M}^2 + {{\Delta \Gamma}^2 \over 4}) t^2
+ 2 Re(\lambda) \Delta\Gamma t + 4 Im(\lambda) \Delta M t \}
\label{mixeqn1}
\end{equation}
up to terms of order $t^2$~\cite{Blaylock}. The decay rate for the
charge conjugate reaction is given by the same expression replacing 
$\lambda$ 
with $\bar{\lambda}$, $B$ with $\bar{B}$, and $q/p$ by $p/q$.
\begin{equation}
\Gamma(\bar{D}^0(t)\to K^-\pi^+) = {e^{-\Gamma t} \over 4} 
|\bar{B}|^2 |{p\over q}|^2
\{ 4 |\bar{\lambda}|^2 + ({\Delta M}^2 + {{\Delta \Gamma}^2 \over 4}) t^2
+ 2 Re(\bar{\lambda}) \Delta\Gamma t + 4 Im(\bar{\lambda}) \Delta M t \}
\label{mixeqn2}
\end{equation}
In the past, it was assumed that the term proportional to $\Delta M ~t$
changes sign when averaging over a sample with equal numbers
of $D^0$ and $\bar{D^0}$ mesons\cite{Bigi},\cite{Browder}. 
This assumption is not correct in general
 as was noted in Reference\cite{Blaylock}.

The previous experimental
analyses\cite{E691}, \cite{Browder}, \cite{E791} 
considered the deterioration
of the limit in the case when the term proportional to 
$\Delta \Gamma ~t$ interfered
destructively with the mixing and DCSD components. 
The Standard Model expectation for $\Delta \Gamma$ is
many orders of magnitude below the current experimental sensitivity so 
this interference scenario is very unlikely. 
In most new physics scenarios which would give $r_{mix}\sim O(10^{-3})$, 
$\Delta\Gamma$ is not enhanced 
whereas values of $\Delta M$ much
larger than those expected from the Standard Model are possible. 
It is also possible to
experimentally verify that $\Delta \Gamma$ can be neglected by measuring
the $D$ meson lifetime in a CP eigenstate e.g. 
$D^0\to K^-K^+$ and comparing to the lifetime
in $D^0\to K^-\pi^+$\cite{liu1},\cite{liu2}.

We now consider equations ~(\ref{mixeqn1}),~(\ref{mixeqn2}) in the following
situation. Let 
\[
{p\over q}=\beta e^{i \phi}
\]
\[
{A\over B}=\alpha e^{i \delta}
\]
and $\displaystyle \alpha^2= 
{\Gamma(D^0\to K^+\pi^-)\over \Gamma(D^0\to K^-\pi^+)}$.
The phase $\phi$ is due to CP violation in the mass matrix.
A non-zero value of $\delta$ may arise if the amplitudes
A and B have different FSI.  Alternatively, if there are
complex contributions to one of the amplitudes (e.g. A) that are not 
present in the other (e.g. B), this can lead to an overall phase in
$\displaystyle{A\over B}$.
We have assumed that there is no
direct CP violation in the amplitudes (and hence e.g.
$\Gamma(D^0\to K^- \pi^+)=\Gamma(\bar{D}^0\to K^+ \pi^-)$)\cite{phase}.
In addition, we neglect the small phase in $A/B$ from the CKM matrix,
which is approximately $A^2\lambda^4\eta$ in the
Wolfenstein parameterization and which lies
in the range $(2.3-5.3)\times 10^{-4}$. 

The decay rate for wrong sign $D^0$ decays to $K^+\pi^-$
is given by
\begin{eqnarray*}
\Gamma(D^0(t)\to K^+\pi^-) &  = &
{e^{-\Gamma t} \over 4} |B|^2
 \times  \\ 
& & {1\over \beta^2} \{ 4 \alpha^2 \beta^2 +
({\Delta M}^2 + {{\Delta \Gamma}^2 \over 4}) t^2  \\
& & + 2 \alpha \beta \cos(\phi+\delta)(\Delta \Gamma t)  
 +4 \sin(\phi+\delta)\alpha\beta \Delta M t  \}
\end{eqnarray*}
The corresponding rate for the charge conjugate reaction is
obtained by replacing $\phi$ the phase from CP violation with $-\phi$
and by changing $\beta$ to $1/\beta$
\begin{eqnarray*}
\Gamma(\bar{D}^0 (t)\to K^-\pi^+) &= & {e^{-\Gamma t} \over 4} 
|\bar{B}|^2 \times \\ 
& & {\beta^2} \{ 4 {\alpha^2 \over \beta^2} +
({\Delta M}^2 + {{\Delta \Gamma}^2 \over 4}) t^2  \\
& & + 2 {\alpha \over \beta} \cos(-\phi+\delta)(\Delta \Gamma t) 
 +4 \sin(-\phi+\delta){\alpha\over\beta} \Delta M t  \}
\end{eqnarray*}

In the experimental analyses, the time dependent rate integrated
over both types of particles is used:
$$
\Gamma (D^0(t)\to K^+\pi^-) + \Gamma (\bar{D}^0(t)\to K^-\pi^+)
$$
This rate, which will be denoted by
$\Gamma(D^0(t)+\bar{D}^0(t))$, is given by
\begin{eqnarray}
\Gamma(D^0(t)+\bar{D}^0(t)) &= & 2 F(t) \times  \\
& & \{ 4 \alpha^2  + 
{1\over 2}(\beta^2+{1\over \beta^2})
({\Delta M}^2 + {{\Delta \Gamma}^2 \over 4}) t^2  \nonumber \\
& &  + \alpha  (\beta\cos(-\phi+\delta)
+{1\over\beta}\cos(\phi+\delta))\Delta \Gamma ~t  \nonumber \\
& & + 2 \alpha (\beta\sin(-\phi+\delta) 
+ {1\over\beta}\sin(\phi+\delta))\Delta M ~t \} \nonumber
\end{eqnarray}
where $\displaystyle F(t)={1\over 4} e^{-\Gamma t} ~|B|^2$.

\bigskip

\section{Effects of FSI and CP Violation}

Two scenarios are considered in what follows. First the
case of no CP violation but significant final state interactions (FSI)
and then the case of both large CP violation and 
significant final state interaction are examined.

\subsection{Effects of FSI}

In the first scenario, consider the case of large mixing
with $\Delta M>>\Delta M_{SM}$,
the value in the Standard Model. 
Assume that this does not lead to an enhancement
of $\Delta \Gamma$ i.e.
$\Delta\Gamma_{SM}=\Delta\Gamma<<\Delta M$ and allow for non-zero $\delta$
but no CP violation ($\phi=0, \beta=1$). The above equation then reduces to
\begin{equation}
\Gamma(D^0(t)+\bar{D}^0(t)) = 2 F(t)
 \{ 4 \alpha^2  +
({\Delta M}^2 + {{\Delta \Gamma}^2 \over 4}) t^2
+4 \alpha (\sin(\delta))\Delta M ~t \}
\end{equation}

In order to determine the size of the new term proportional 
to $\Delta M ~t$,
the values of the phase difference $\delta$ are considered
in various models. This will allow
an estimate of the additional 
experimental systematic error that is incurred from ignoring FSI.
This phase difference $\delta$ is zero in the limit of exact SU(3) symmetry. 
The values of $\delta$ from various models
are given in Table~\ref{Tbdelta}. 
Large values of the phase $\delta$ occur when 
SU(3) breaking is largest. We 
use the experimental result from CLEO~II for $D^0\to K^+ \pi^-$
($\alpha^2=0.0077\pm 0.0025 \pm 0.0025$) and assume that it is entirely
due to DCSD. This is found numerically 
to give the most conservative upper limit
on the size of the interference effect.

In general, the amplitudes for the $D^0\to K^- \pi^+$ 
and $D^0\to K^+ \pi^-$ can be written as:
\begin{equation}
A(D^0\to K^-\pi^+) = e^{i \delta_3} 
[(A_1 + C) e^{i (\delta_1-\delta_3)} + A_3 ]
\end{equation}
\begin{equation}
A(D^0\to K^+\pi^-) = -\theta_c^2
e^{i \delta_3} [ (\tilde{A}_1 + \tilde{C}) e^{i (\delta_1-\delta_3)} 
+ \tilde{A}_3 ]
\end{equation}
where $A_1$ and $A_3$ are the quark decay contributions into $I=1/2$ and
$I=3/2$ final states respectively. C is the W-exchange contribution and
$\delta_1$ and $\delta_3$ are the FSI phases. $\tilde{A_i}$, $\tilde{C}$
are the corresponding DCSD amplitudes after the CKM factor $-\theta_c^2$
has been factored out. 
The phase shifts in a given isospin eigenstate for particles
and antiparticles are identical by CPT invariance (which we assume
as stated explicitly). 

The phase $\delta$ vanishes if two conditions are satisfied:
(i) $\delta_1-\delta_3= \tilde{\delta}_1-\tilde{\delta}_3$ 
and (ii) $A_3/A_1=\tilde{A_3}/\tilde{A_1}$. The first condition
follows from CPT invariance and the second is satisfied if $SU(3)$
symmetry holds. Hence if $SU(3)$ is an approximate symmetry, the phase
$\delta$ should be small. The models used have been tuned to reproduce
the observed magnitude of $SU(3)$ breaking in $D$ decays.
To obtain more information, we turn to the detailed
model fits. 

\subsection{Details of the Models}

In the model of Chau and Cheng,
\begin{equation}
A_1 \cong 0.82, ~A_3\cong 0.16,~C\cong -0.13
\end{equation}
\begin{equation}
\tilde{A}_1 \cong 1.14, ~\tilde{A}_3\cong 0.33, ~\tilde{C}\approx C
\end{equation}
and
\begin{equation}
\delta_1-\delta_3 \approx 90^{0},~ \delta_3\approx 0. 
\end{equation}
Then 
\begin{equation}
A(D^0\to K^-\pi^+)\cong (0.72) e^{i ~76^{0}}
\end{equation}
\begin{equation}
A(D^0\to K^+\pi^-)\cong -\theta_c^2 (1.01) e^{i ~72^{0}}
\end{equation}
This yields a phase difference between the two decay modes
of $\delta=4^{0}$. If the W-exchange
contribution $C$ is omitted, the phase difference becomes
$\delta=5^{0}$.

In the model of Buccella {\it et al.}, one has
\begin{equation}
{A}_1 \cong 4.35,~ {A}_3\cong -2.3, ~{C}\approx -0.5
\end{equation}
\begin{equation}
\tilde{A}_1 \cong 5.2, ~\tilde{A}_3\cong -2.3, ~\tilde{C}\approx -C
\end{equation}
and
\begin{equation}
\delta_1 - \delta_3 \cong 25^{0}, \delta_3\approx 0.
\end{equation}
Then
\begin{equation}
A(D^0\to K^-\pi^+)\cong (2) e^{i ~54.3^{0}}
\end{equation}
\begin{equation}
A(D^0\to K^+\pi^-)\cong -\theta_c^2 (3.7) e^{i ~41.2^{0}}
\end{equation}
leading to a phase difference of $\delta=13^{0}$. Omitting the
W-exchange term gives a slightly smaller value of $6^{0}$.  
It should be noted that a $\delta_1=25^{0}$ relates to 
$\delta_R$ for the $I=1/2$ $0^+$ resonance in the $K\pi$ channel
by 
\begin{equation}
\tan \delta_R = {\Gamma \over {2 \Delta}} = {B \sin\delta_1 \over
{B \cos\delta_1 + (1- B)}}
\end{equation}
where $B=BR(0^+\to K \pi)\approx 0.50$, $\Gamma\approx 200$ MeV,
$\Delta=M_R- M_D\approx 70$ MeV and $\delta_R = (55-65)^{0}$.

The models discussed above predict 
\begin{equation}
{ {BR(D^0\to K^+\pi^-)} \over { BR(D^0\to K^-\pi^+)}} = 
(2.3-3.4)\tan^4\theta_c
\end{equation}
which is compatible with the CLEO~II measurement. 
There are also other models for $D$ decays in which a value for
the phase difference $\delta$ can be extracted\cite{Kaedin}. 
Since it is difficult
to assign errors to these predictions, we regard $0^{0}-13^{0}$ as
a reasonable range for $\delta$. 
In order to explore the range of $\delta$ in the models, we have
calculated the value of $\delta$ omitting the W-exchange term.
This corresponds to a dramatic change in the parameters of the models.

\begin{table}[htb]
\caption{Values of $\delta$ in various phenomenological
models of $D$ meson decay.}
\label{Tbdelta}
\begin{tabular}{ll}
          & $\delta$ \\ \hline 
Exact SU(3) limit \cite{wolf2} & $0^{0}$ \\
Chau and Cheng\cite{chaucheng} & $4^{0}$ \\
Chau and Cheng (no W-exchange)\cite{chaucheng} & $5^{0}$ \\
Buccella et al.  & $13^{0}$ \\
Buccella et al. (no W-exchange) & $6^{0}$ \\
\end{tabular}
\end{table}

\subsection{Summary of the Interference Effect from FSI}

To summarize, the phenomenological models which have been tuned
to agree with the observed branching fractions
and the fits to the $D$ meson data give $\delta$
in the range of $5^{0}-13^{0}$. To evaluate the possible experimental
consequences,  consider the case of maximal
destructive interference ($\phi=0$, $\beta=1$), with $\delta=13^{0}$. 
We allow a one standard deviation variation on
$R_{DCSD}=\alpha^2$ from
the CLEO~II measurement in order to obtain an upper limit on the effect
of the interference term.
We set $r_{mix}$, the ratio of integrated rates for mixed events relative to
unmixed events, to the E691 upper bound\cite{defrmix}.   
The contributions
of the mixing term, the DCSD term, and the term proportional
to $\Delta M t$ are shown in Figure 1. These time dependent searches
are most sensitive to excess events from
mixing for $t>0.22$ ps $\displaystyle ={\tau_{D^0}\over 2}$, where the 
combinatorial backgrounds are manageable and where the mixing term
is expected to peak. In addition, there is no loss in efficiency
for the mixing component when this cut is imposed. An upper limit
of $t<4.0$ ps is also imposed. 
The change in the observed event yield for various values 
of $R_{DCSD}$ and maximal destructive interference
 are given in Table~\ref{Tbyield}. These were
calculated for the scenario
with maximal destructive interference and $\delta=13^{0}$.
We also give the change in the observed event yield for $t>2\tau_{D^0}$ 
(this is the region where mixing peaks and the
experiments are most sensitive) in Table~\ref{Tblong}. 
This change is at most 10-15\% and is well within the experimental
systematic error assigned by the E691 and E791 experiments to their limits.

\begin{table}[htb]
\caption{The change in wrong sign event yield 
for $t>0.22$ ps with maximal destructive interference, $r_{mix}=0.37\%$,
and $\delta=13^{0}$.}
\medskip
\label{Tbyield}
\begin{tabular}{ll}
  $R_{DCSD}$        & $\Delta$ Yield (\%) \\ \hline 
   0.0052 & $ 10\%$ \\
   0.0077 & $ 8\%$ \\
   0.0102 & $ 1\%$ \\
\end{tabular}
\end{table}

\begin{table}[htb]
\caption{The change in wrong sign event yield 
for $t>0.88$ ps (2$\tau_{D^0}$) with maximal destructive interference,
$r_{mix}=0.37\%$, and $\delta=13^{0}$.}
\medskip
\label{Tblong}
\begin{tabular}{ll}
  $R_{DCSD}$        & $\Delta$ Yield (\%) \\ \hline 
   0.0052 & $ 12\%$ \\
   0.0077 & $ 10\%$ \\
   0.0102 & $ 9\%$ \\
\end{tabular}
\end{table}

\section{Effects of CP Violation}

Now consider the contribution of CP violation. Let
$\beta=1 -\epsilon$ and 
$\displaystyle{1\over \beta}=1+\epsilon$\cite{epsfoot}. 
We assume $\epsilon$ is small compared to 1 and retain only terms linear
in $\epsilon$; this is justified in the SM and even more so when
$\Delta M$ is enhanced and $\Delta \Gamma/\Delta M<<1$. We allow the phase 
$\phi$ to be arbitrary. 
With these definitions and $\Delta \Gamma < <\Delta M$, the expression
for $\Gamma(D^0(t)+\bar{D}^0(t))$ now becomes:
\begin{eqnarray}
\Gamma(D^0(t)+\bar{D}^0(t)) & = &  \\
2 F(t) \times &   &  \{ 4 \alpha^2  +
({\Delta M}^2 + {{\Delta \Gamma}^2 \over 4}) t^2 \nonumber \\
& & +  \alpha  (\cos(-\phi+\delta)+\cos(\phi+\delta))\Delta \Gamma ~t 
 +\alpha \epsilon (\cos(\phi+\delta)-
\cos(-\phi+\delta))\Delta \Gamma ~t \nonumber \\
& & +2 \alpha (\sin(-\phi+\delta) + \sin(\phi+\delta))\Delta M ~t 
 +2 \alpha \epsilon(\sin(-\phi+\delta) 
- \sin(\phi+\delta))\Delta M ~t \} \nonumber
\end{eqnarray}

The quantity
$\epsilon$ is assumed to be small as in Ref\cite{Blaylock}, however,
the CP violating phase $\phi$ can be large as is the
case for certain extensions of the Standard Model.
The quantity $\epsilon$ for $D$ mixing is given by\cite{BGHP}
\begin{equation}
\epsilon \approx {-2 ~{\rm Im}({M_{12}^* \Gamma_{12} \over 2}) \over 
{ {1\over 2} \Delta M^2 + {\Delta \Gamma^2\over 4} } }
\end{equation}
in the Standard Model and is already small ($\epsilon<O(2\%)$)\cite{BGHP}.

In new physics scenarios with $\Delta\Gamma_{SM}=\Delta\Gamma<<\Delta M$,
\begin{equation}
\tan\phi \cong  {{\rm Im}(M_{12}) \over \Delta M} 
\end{equation}
For non standard models with ${\rm Im}(M_{12})/\Delta M$ 
of order unity,  $\tan\phi$ may be large (O(1)).
By contrast, 
\begin{equation}
{\epsilon} \cong 2 ( {\Delta \Gamma \over \Delta M}) 
{{\rm Im}(M_{12})\over \Delta M} 
\approx 2({\Delta\Gamma \over \Delta M}) << 1
\end{equation}
The crucial point is that $\epsilon$ is proportional to 
$1/\Delta M$ and is highly suppressed if $\tan\phi$ is of order unity
and $\Delta M$ is enhanced.
It is important to note that while $\tan(\phi)$ can be much larger than
the Standard Model expectation
 $\epsilon$ will be even smaller than
the value in the Standard Model for new physics scenarios in which
$\Delta M$ is enhanced.

The total wrong sign rate can then be reduced to 
\begin{eqnarray}
\Gamma(D^0(t)+\bar{D}^0(t)) & = &  2 F(t) \\
& &  \{ 4 \alpha^2  +
({\Delta M}^2 + {{\Delta \Gamma}^2 \over 4}) t^2 \nonumber \\
& & +  2 \alpha  (\cos(\phi)\cos\delta))\Delta \Gamma ~t 
 - 2 \alpha \epsilon (\sin(\phi)\sin(\delta))\Delta \Gamma ~t \nonumber \\
& & +4 \alpha \epsilon (\cos(\delta)\sin(\phi))\Delta M ~t 
 +4 \alpha (\sin(\delta) \cos(\phi))\Delta M ~t \} \nonumber
\end{eqnarray}
With $\epsilon$ as given above and $\Delta \Gamma << \Delta M$,
the expression for $\Gamma(D^0(t)+\bar{D}^0(t))$ becomes
\begin{eqnarray*}
\Gamma(D^0(t)+\bar{D}^0(t)) & = &  2 F(t) \\
& &  \{ 4 \alpha^2  +
({\Delta M}^2 ) t^2  \nonumber \\
& & 
 +4 \alpha (\sin(\delta) \cos(\phi))\Delta M ~t \} \nonumber
\end{eqnarray*}
Hence, the term due to CP violation is too small to be observable
when $\Delta M$ and $\rm{Im}(M_{12})$ are enhanced.

As experimental sensitivity improves and become
sensitive to mixing at the level $r_{mix}<10^{-4}$, it is possible
that better sensitivity to $D^0-\bar{D}^0$ mixing can
be achieved by fitting the time distribution of
$\Gamma(D^0\to K^+\pi^-)-\Gamma(\bar{D}^0\to K^-\pi^+)$.
This rate, which will henceforth
be denoted $\Gamma(D^0(t)-\bar{D}^0(t))$, is given by
\begin{eqnarray}
\Gamma(D^0(t)-\bar{D}^0(t)) & = & 2 F(t) \times \\
&& \{ 
 2 \alpha \epsilon (\cos(\phi)\cos\delta))\Delta \Gamma ~t
- 2 \alpha  (\sin(\phi)\sin(\delta))\Delta \Gamma ~t \nonumber \\
&& +4 \alpha (\cos(\delta)\sin(\phi))\Delta M ~t
+4 \alpha \epsilon (\sin(\delta) \cos(\phi))\Delta M ~t \} \nonumber
\end{eqnarray}
In the limit that $\Delta\Gamma<< \Delta M$ and $\phi$ is large,
this reduces to
\begin{equation}
\Gamma(D^0(t)-\bar{D}^0(t)) \cong 2 F(t)
 \{
 4 \alpha (\cos(\delta)\sin(\phi))\Delta M ~t
+4 \alpha \epsilon (\sin(\delta) \sin(\phi))\Delta M ~t \}
\end{equation}
or neglecting the small term proportional to $\epsilon \sin(\delta)$,
\begin{equation}
\Gamma(D^0(t)-\bar{D}^0(t)) \cong
2 F(t) [4 \alpha (\cos(\delta)\sin(\phi))\Delta M ~t ]
\end{equation}
Note that in this case, the long lived tail of DCSD does not
contribute to the signal. 
In addition, as noted by Wolfenstein\cite{wolf2},
for small values of $\Delta M$, the term proportional
to $\Delta M ~t$ will be larger than the
term in $\Gamma(D^0(t)+\bar{D}^0(t))$ which is
proportional to $(\Delta M ~t)^2$. This feature is illustrated
in Figs. 2~(a),~2~(b).

\section{Conclusions}

The formalism presented here must be modified 
for the case of multibody  modes such as $D^0\to K^+ \pi^- \pi^0$
or $D^0\to K^+ \pi^- \pi^+ \pi^-$. For these other modes, 
an additional complication is that
the value of the final state phase difference, $\delta$,
may be different from the value in the case of $D^0\to K^+ \pi^-/
\bar{D}^0\to K^+\pi^-$
and is not guaranteed to be small.
It should also be remembered that limits on $D^0-\bar{D}^0$ mixing
from studies of
semileptonic decays do not have the complications from DCSD and
other hadronic effects discussed here.

At the present level
of sensitivity and with reasonable (though model dependent) values
for the phase difference $\delta$, 
the $\Delta M~t$ term which arises 
from FSI does not dramatically change the observed event yield for
experiments which study the time dependence of mixing and is not yet a 
significant systematic experimental limitation. 
We suggest that future experiments determine systematic
errors on their limits by using an upper limit on the phase difference
$\delta$. 

The contribution
from the corresponding term proportional to $\Delta M~t$ due to
CP violation which arises 
in extensions of Standard Model is highly suppressed. This 
term is not observable at the present level of experimental sensitivity.
However, as emphasized by Liu\cite{liu1} 
and by Wolfenstein\cite{wolf2}, this term should not
be neglected as
experimental examination of the $D^0(t)-\bar{D}^0(t)$ distribution
may allow more sensitive searches for $D^0-\bar{D^0}$ mixing 
in the future if the CP violating phase is large.

This work was supported in part by the United States
Department of Energy under grant DE-FG 03-94ER40833 and by Tokkuri Tei. 
We thank 
G. Burdman, E. Golowich, J. Hewett, D. Kaplan, T. Liu, Y. Nir,
and M. Witherell for useful and enjoyable
discussions.

\pagebreak

\bigskip
\begin{figure}[htb]
\begin{center}
\unitlength 1.0in
\begin{picture}(3.,3.0)(0,0)
\put(-1.00,-1.50)
{\psfig{bbllx=0pt,bblly=0pt,width=3.0in,height=3.0in,file=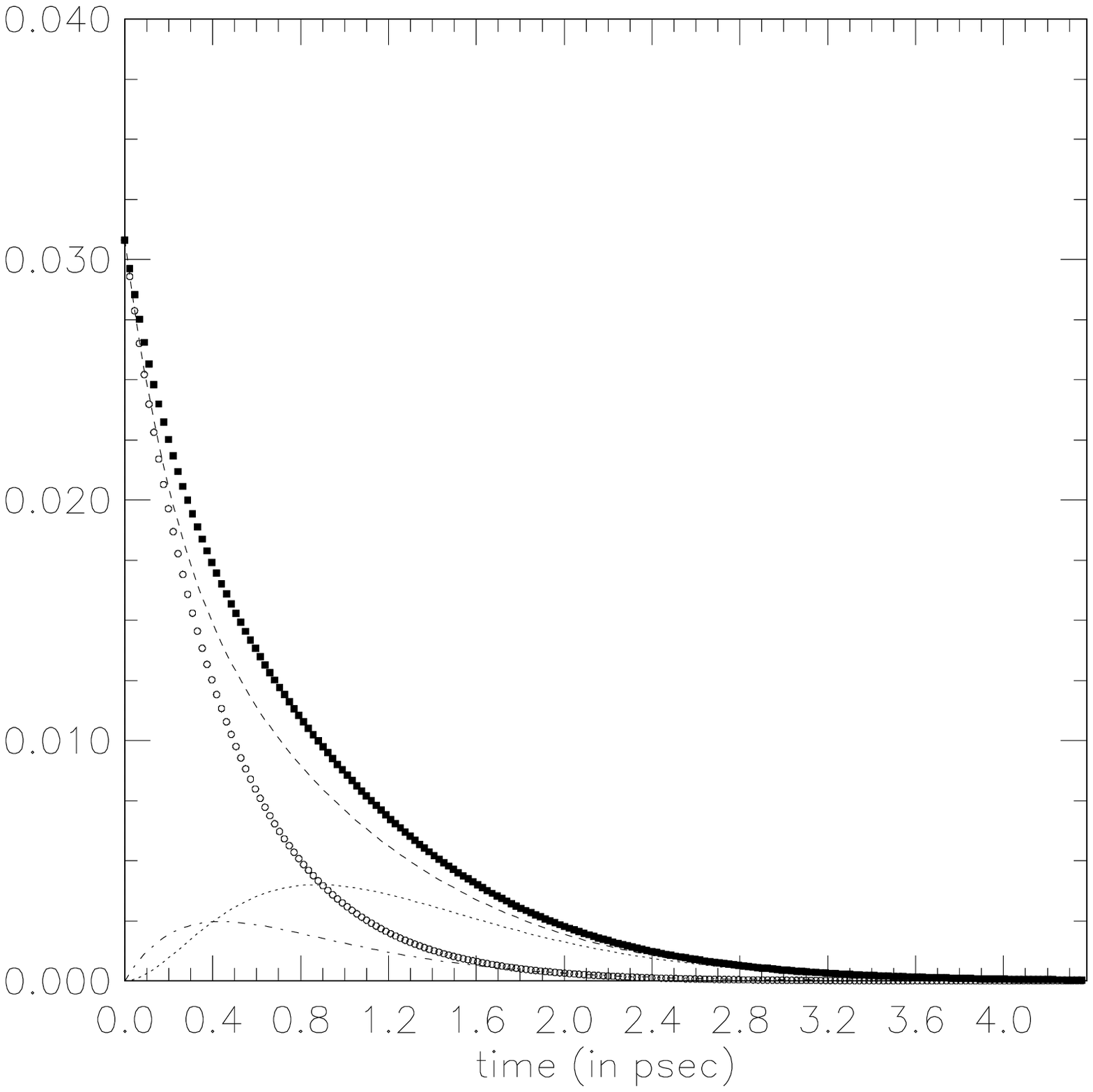}}
\end{picture}
\vskip 80 mm
\caption{The case of maximal destructive interference.
The time dependence of DCSD(open circles), 
mixing(dotted), interference(dash-dotted)
mixing and DCSD without
interference(solid points), mixing and DCSD with interference (dashed) 
are shown.
For this plot,
the mixing rate is taken to be $\Delta M/\Gamma=0.086$ ($r_{mix}=0.37\%$)
and the DCSD rate is taken to be the central value of
the branching fraction for $D^0\to K^+\pi^-$ determined by CLEO~II (i.e.
$R_{DCSD}=0.0077$). Other scenarios are discussed in the text.}
\label{fig1}
\end{center}
\end{figure}
\bigskip

\bigskip
\begin{figure}[htb]
\begin{center}
\unitlength 1.0in
\begin{picture}(3.,3.0)(0,0)
\put(-1.00,-1.50)
{\psfig{bbllx=0pt,bblly=0pt,width=3.0in,height=3.0in,file=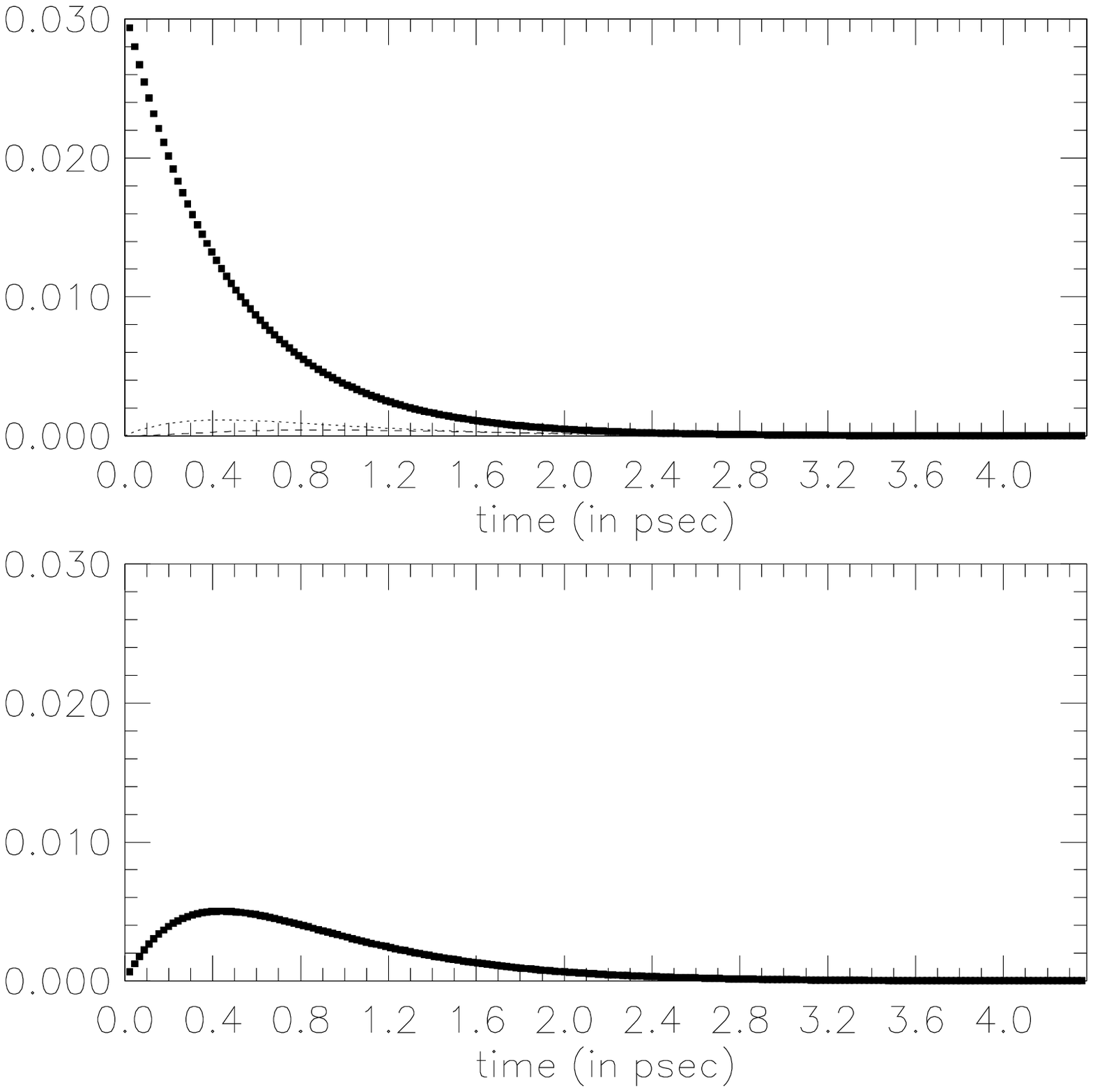}}
\end{picture}
\vskip 80 mm
\caption{The time dependence for wrong sign events 
for the case of small mixing ($\Delta M/\Gamma=0.02$)
for (a) $\Gamma(D^0(t)+\bar{D}^0(t))$, where
the dotted component is the FSI interference
term and the dashed component is the usual mixing term (both scaled
up by a factor of two in order to be visible),
 and for (b) $\Gamma(D^0(t)-\bar{D}^0(t))$. 
Note that in (b) 
there is no background to mixing from DCSD.}
\label{fig2}
\end{center}
\end{figure}
\bigskip


\begin{references}
\bibitem{phase}{Strictly speaking, the phase $\phi$ is not observable
since it depends on the phase convention. However, the combinations
of $\phi$ with the phases in the amplitudes of $D^0\to K^- \pi^+$
and $D^0\to K^+\pi^-$ are observable. With the usual convention,
these phases are negligible and hence $\phi$ is effectively observable.
We are assuming that contributions from new physics preserve this
feature.}
\bibitem{defrmix}{The ratio of integrated rates for mixed and
unmixed events is $${{(\Delta M/\Gamma)^2 + (\Delta\Gamma/\Gamma)^2/4}
\over{2 +(\Delta M/\Gamma)^2}}.$$ In the limit that the width
difference is small, $r_{mix}=1/2 (\Delta M/\Gamma)^2$.}
\bibitem{epsfoot}{For small values, $\epsilon$ is related to 
the usual parameter $\epsilon_{D}$ by 
$\epsilon \approx -2 \rm{Re}(\epsilon_D)$.}
\end{references}
\end{document}